\documentclass{emulateapj}

\usepackage{graphicx}
\usepackage{epstopdf}

\newcommand{\lx}{erg~s$^{-1}$}
\newcommand{\nh}{cm$^{-2}$}
\newcommand{\chandra}{{\it Chandra}}
\newcommand{\xmm}{{\it XMM-Newton}}

\newcommand{\msun}{$M_{\odot}$}

\newcommand{\fxfv}{$log_{10}\frac{f_x}{f_v}$}

\shorttitle{IC 10 X-ray Variability}
\shortauthors{Laycock et al.}

\begin{document}


\title{The X-ray Binary Population of the Nearby Dwarf Starburst Galaxy IC 10: Variable and Transient X-ray Sources }


\author{Silas G. T. Laycock \altaffilmark{1,2,*}, Rigel Cappallo\altaffilmark{1,2}, Benjamin F. Williams \altaffilmark{3}, Andrea Prestwich\altaffilmark{4}, Breanna Binder\altaffilmark{3,5} Dimitris M. Christodoulou\altaffilmark{1,6}}
\altaffiltext{1}{Lowell Center for Space Science and Technology, 600 Suffolk Street, Lowell, MA 01854, USA}
\altaffiltext{2}{Department of Physics and Applied Physics, University of Massachusetts Lowell, MA, 01854 }
\altaffiltext{3}{University of Washington, Department of Astronomy, Box 351580, Seattle, WA 98195}
\altaffiltext{4}{Harvard-Smithsonian Center for Astrophysics, 60 Garden St, Cambridge, MA, 02138, USA}
\altaffiltext{5}{Department of Physics \& Astronomy, California State Polytechnic University, 3801 West Temple Ave, Pomona, CA 91768}
\altaffiltext{6}{Department of Mathematical Sciences, University of Massachusetts Lowell, MA, 01854 }
\altaffiltext{*}{Email: silas\_laycock@uml.edu }


\begin{abstract} 
We have monitored the Cassiopeia dwarf galaxy (IC 10) in a series of 10 \chandra~ ACIS-S observations to capture its variable and transient X-ray source population, which is expected to be dominated by High Mass X-ray Binaries (HMXBs). We present a sample of 21 X-ray sources that are variable between observations at the 3$\sigma$ level, from a catalog of 110 unique point sources. 
We find 4 transients  (flux variability ratio greater than 10) and a further 8 objects with ratio $>$ 5. The observations span years 2003 - 2010 and reach a limiting luminosity of $>10^{35}$\lx, providing sensitivity to X-ray binaries in IC 10 as well as flare stars in the foreground Milky Way. The nature of the variable sources is investigated from light-curves, X-ray spectra, energy quantiles, and optical counterparts.  The purpose of this study is to discover the composition of the X-ray binary population in a young starburst environment. IC 10 provides a sharp contrast in stellar population age ($<$10 My) when compared to the Magellanic Clouds (40-200 My) where most of the known HMXBs reside.  We find 10 strong HMXB candidates, 2 probable background Active Galactic Nuclei, 4 foreground flare-stars or active binaries, and 5 not yet classifiable sources. Complete classification of the sample requires optical spectroscopy for radial velocity analysis and deeper X-ray observations to obtain higher S/N spectra and search for pulsations. A catalog and supporting dataset are provided.

\end {abstract}


\section{Introduction}

The dwarf-irregular galaxy IC 10 in Cassiopeia presents one of the most active starbursts in the local group and a unique environment to study the immediate relics of the most massive stars: black holes and neutron stars in high mass X-ray binaries (HMXBs). Motivated by the discovery of the black hole HMXB IC 10 X-1 \citep{Prestwich2007} and the very large amplitude transient IC 10 X-2 \citep{Laycock2014} we conducted a 7-shot monitoring campaign with \chandra ~during 2009-10. Combined with deep observations taken in 2003 and 2006, we were able to span timescales ranging up to 7 years reaching a limiting sensitivity of $>10^{35}$\lx.  This cadence and sensitivity, coupled with the angular resolution of \chandra, enable light-curves to be constructed for individual point sources.

IC 10 hosts a young ($\leq$6 $\times 10^{6}$ yr) stellar population \citep{Massey2007} accompanied by the highest known space-density of Wolf Rayet (WR) stars \citep{Crowther2003}, which are the evolved helium cores of stars with initial stellar masses in excess of 20 \msun.  The star formation rate (SFR) is reported to be as high as 0.5 \msun $yr^{-1}$ \citep{Leroy2006, Massey2002}, which when normalized by the mass of the galaxy (2 $\times 10^{7}$ \msun;  \cite{Petitpas1997}) yields 2.5 $\times 10^{-8}$ \msun $yr^{-1}$ \msun$^{-1}$ which is among the highest specific SFRs in the local universe exceeding the values for the Milky Way (MW), Small Magellanic Cloud (SMC), and Large Magellanic Cloud (LMC).  Depending on the indicators and methods used, there is an order of magnitude dispersion among SFR values for IC 10 \citep{Chomiuk2011, Heesen2007, Leroy2006}. The chaotic structure of IC 10 is probably part of the picture and indeed much of the galaxy resembles an OB association, but on a kiloparsec scale \citep{Massey2002}. IC 10 is replete with a complex structure of $H\alpha$-bright and radio-bright bubbles, filaments, and HII regions, all signaling its turbulent evolution under the influence of the kinetic energy being injected by winds and supernovae.

The similarities between IC10 and the SMC are striking; both are gas-rich dwarf irregulars and both are thought to have undergone tidal disruption in the recent past \citep{Besla2012, Nidever2013} leading to triggered star formation activity. However, two critical differences exist between these galaxies that make IC 10 an exciting new laboratory for stellar astrophysics.  The first and most important contrast is the time since star formation began. This is $\leq$6 $\times 10^{6}$ yr for IC 10 compared to 40-200 $\times 10^{6}$ yr for the various distinct age-segregated populations identified in the SMC \citep{Zaritsky2004}, where the HMXBs are found in the 40-70 Myr sub-population \citep{Antoniou2010}. The second contrast is in metallicity which for IC 10 is  $Z\simeq Z_{\odot}/5$  \citep{Garnett1990},  intermediate between the SMC and the MW.  Complete censuses of X-ray binary populations in local group galaxies are a powerful approach to discover the underlying physics of stellar evolution, for example the effect of metallicity. The Magellanic clouds have historically served this purpose, but gaining proper control of secular differences between galaxies requires new independent laboratories such as IC 10 to be added. Important large samples have been constructed by mining the \chandra ~archive, (for example \citealt{Fabbiano2006, Brorby2014,Prestwich2013}) and by concerted multi-year efforts to intensively observe nearby galaxies (e.g., \citealt{Tullman2011, Binder2015, Williams2015, Li2010}). An X-ray study of point sources in IC 10 was conducted by \cite{Wang2005} using single epoch data from \xmm ~and \chandra. Among their important findings was a population of point sources significantly above the expected background density and spatially concentrated within the optical outline of the galaxy. The combined X-ray spectrum of these point-sources was consistent with an absorbed power-law with parameters characteristic of HMXBs.  

 \begin{figure*}
\begin{center}
\includegraphics[angle=0,width=16cm, trim={2cm 3.5cm 0 3cm}]{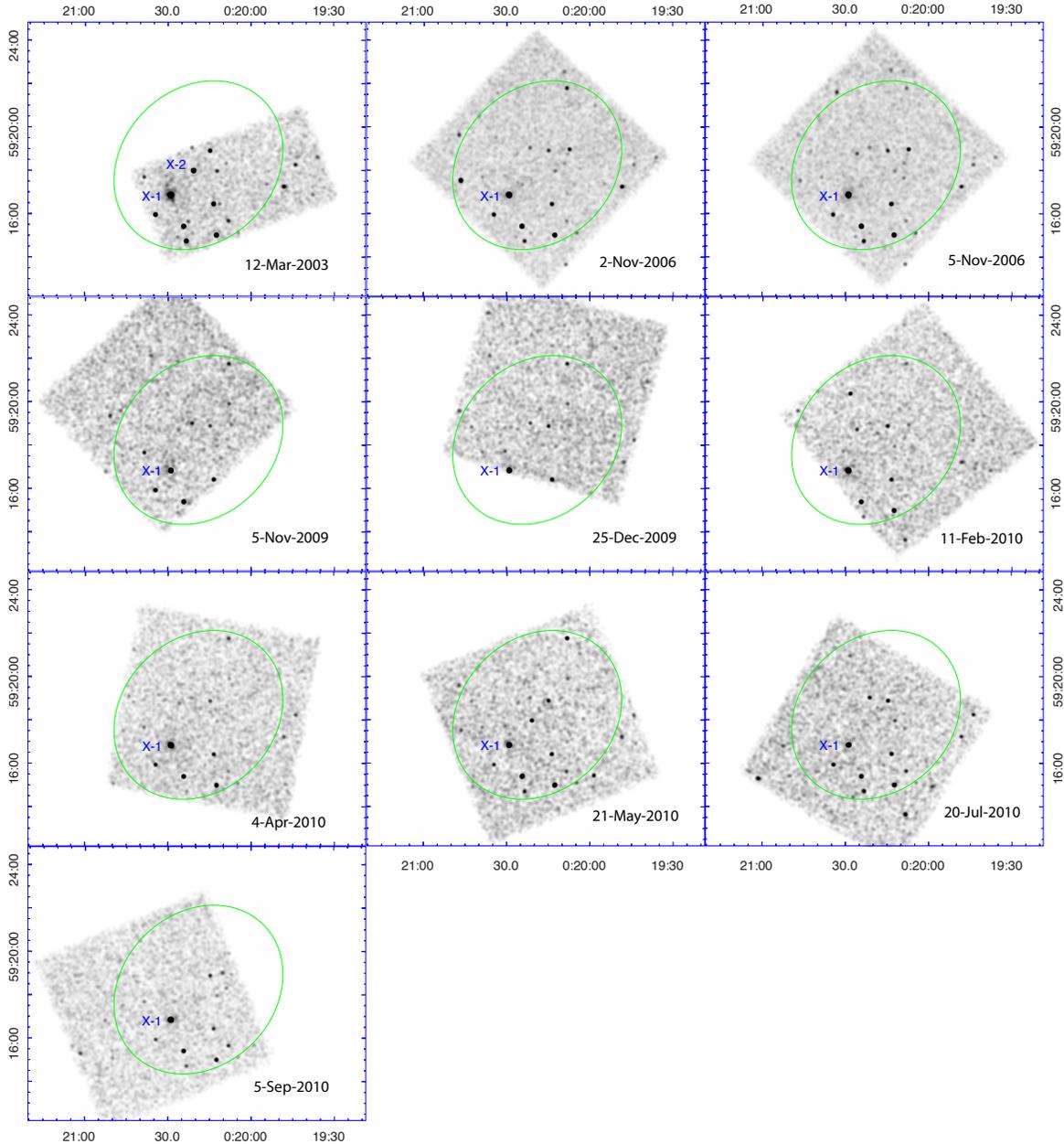}
\caption{\chandra ~X-ray images of IC 10 using the ACIS-S3 CCD sensor. Images are shown in chronological order at identical scale and orientation to show the field of view and prominent point sources in each observation. Green oval delineates the approximate optical extent of the galaxy.  Images are smoothed with a gaussian kernel (width 15 pixels). The different aim-point locations and exposure times lead to some variation in appearance. The coordinate axes are RA, Dec (J2000), see Table~\ref{tab:dataset} for observation parameters.}
\label{fig:cxo_images}
\end{center}
\end{figure*}

Key physics goals to be addressed by IC 10's X-ray binary (XRB) sample include: constructing the X-ray luminosity function (XLF) in order to compare it with the SFR history; discovering what type of XRBs are formed at the earliest times; and breaking the degeneracy between compact-object mass and accretion-mechanism that currently bedevils attempts to understand the nature of both ultraluminous X-ray (ULX) sources and the XLF-SFR relation. The relative proximity of IC10 is crucial because the individual XRBs can be studied in detail once they are identified. In the larger samples of galaxies, the questions are addressed on a purely statistical basis.  Identifying optical counterparts to X-ray sources at the 660 kpc distance of IC 10 is a daunting prospect since it lies at low galactic latitude and it is viewed by us through the outer portion of the Galactic Plane. Since most XRBs are highly variable, a monitoring campaign offers an alternative path to discovering the nature of these individual sources. 

This paper begins with a uniform reduction/analysis of the \chandra ~monitoring and archival observations and visualizations of the CCD data (\S~\ref{sect:obs}) followed by the construction of the source catalog (\S~\ref{sect:catalog}) and the generation of X-ray light curves (\S~\ref{sect:lc}). We outline our spectral analysis using quantiles and CCD pulse-height spectra in  \S~\ref{sect:quantile}, we briefly describe how optical counterparts were identified (\S~\ref{sect:optical}), and then we proceed to \S~\ref{sect:discussion}, where we look at each variable source in turn in order to summarize the available information and to discuss its likely nature. Finally, in \S~\ref{sect:conclusions}, we present an overview of the results and we lay out the direction of future progress. 

\section{Observations}
\label{sect:obs}
A monitoring series of 7$\times$15 ksec \chandra ~observations, spaced at roughly 6-week intervals was obtained during 2009-2010. We used the Advanced CCD Imaging Spectrometer (ACIS), positioning the galaxy IC 10 on the back-illuminated chip S3.  The observing cadence was designed to match the native variability timescales of HMXBs which tend to be transients with typical outbursts lasting for 1-3 weeks repeating on intervals of months. The individual exposure times were calculated to reach a limiting luminosity of 10$^{35}$ \lx ~(assuming 5 net ACIS-S counts at the High Resolution Mirror Assembly (HRMA) aim-point for a detection threshold). This sensitivity would reach the typical luminosity of HMXBs (both persistent supergiant HMXBs and Be-HMXBs in outburst). A pair of very deep ACIS-S3 observations (2$\times$45 ksec) made in November 2006  provide a reference dataset for improved source positions and spectral information. The original \cite{Wang2005} \chandra ~(ACIS-S3, 1/2 subarray) observation of 30 ksec made in 2003 was included in our analysis to provide additional time baseline for the variability study and to make use of their published source identifications.  The complete listing of 10 observation identifiers (ObsIDs) comprising the dataset is summarized in Table~\ref{tab:dataset}. All of the data were processed in a uniform way following the standard {\it Ciao}\footnote{http://cxc.harvard.edu/ciao} threads, and the prescription of \cite{Hong2005} for consistent selection of energy bands---B (0.3-8 keV), S (0.3-1.5 keV), and H (2.5-8 keV)---and extraction regions.
 
\begin{deluxetable*}{llllllll}
\tablecaption{\chandra ~Observations of IC10
 \label{tab:dataset}} 
\tablehead{ 
\colhead{MJD}  &   \colhead{Date} &  \colhead{ObsID}  & \colhead{Exposure}             & \colhead{R.A.}   & \colhead{Dec.} & \colhead{Roll Angle}  & \colhead{$N_{Src}$}  \\
                         &                            &                              &  \colhead{\scriptsize ksec}   &   \colhead{hh:mm:ss}    &  \colhead{dd:mm:ss}    &    \colhead{degrees}     &                                \\      }
\startdata 
52710.7            & 12 Mar 2003      & 03953(a)                   & 28.9                                    &    00:20:25  &  59:16:55      &    339.27                    &     31                        \\
54041.8           & 2 Nov 2006        & 07082                   & 40.1                                   &    00:20:04      &   59:16:45     &    223.70                      &       48                     \\
54044.2            & 5 Nov 2006        & 08458                   & 40.5                                  &    00:20:04     & 59:16:45         &  223.70                        &        41                     \\
55140.7            &  5 Nov 2009       & 11080                   & 14.6                                  &     00:20:17      &  59:17:56       &  226.53               &      19                      \\
55190.2            &  25 Dec 2009     & 11081                   & 8.1                                    &     00:20:19 & 59:18:02    &    286.15      &       24                            \\
55238.5           & 11 Feb 2010      & 11082                   & 14.7                                   &   00:20:23    &  59:17:10  &    320.56       &       24                       \\
 55290.6           & 4 Apr 2010        & 11083                    & 14.7                                 &  00:20:34  &  59:19:01    &      10.32      &        25                       \\
55337.8            & 21 May 2010     & 11084                    & 14.2                                 &  00:20:25    &   59:20:16   &     67.89        &         27                     \\
55397.5            & 20 Jul 2010       & 11085                    & 14.5                                 &  00:20:11  &   59:19:13    &    121.25         &        22                       \\
55444.6            & 5 Sep 2010       & 11086                    & 14.7                                 &   00:20:15  &  59:18:11    &    157.71   &         27                     \\
(b)			&  2-5 Nov 2009   &  57082  		   & 80.6 				     &	    -			&   	-		  &  		-	     &      63		\\		
	
\enddata
\tablecomments{Instrument is the \chandra ~ACIS-S3 CCD. Aim-point coordinates and spacecraft Roll Angle are listed for each observation. $N_{Src}$ is the number of unique point sources detected in each observation after combining {\it wavdetect} lists from the S (0.3-1.5 keV), B (0.3-8 keV), and H (2.5-8 keV) energy bands.\\
(a) {\it Obsid 03953} used about half of the CCD area in sub-array mode.\\
(b) Merged 2006 dataset referred to as {\it obsid 57082} consists of the nearly contiguous {\it ObsIDs 07082 \& 08458} which had identical pointing. }
\end{deluxetable*}

 \subsection{Images}
 
\begin{figure*}
\begin{center}
\includegraphics[angle=0,width=16cm]{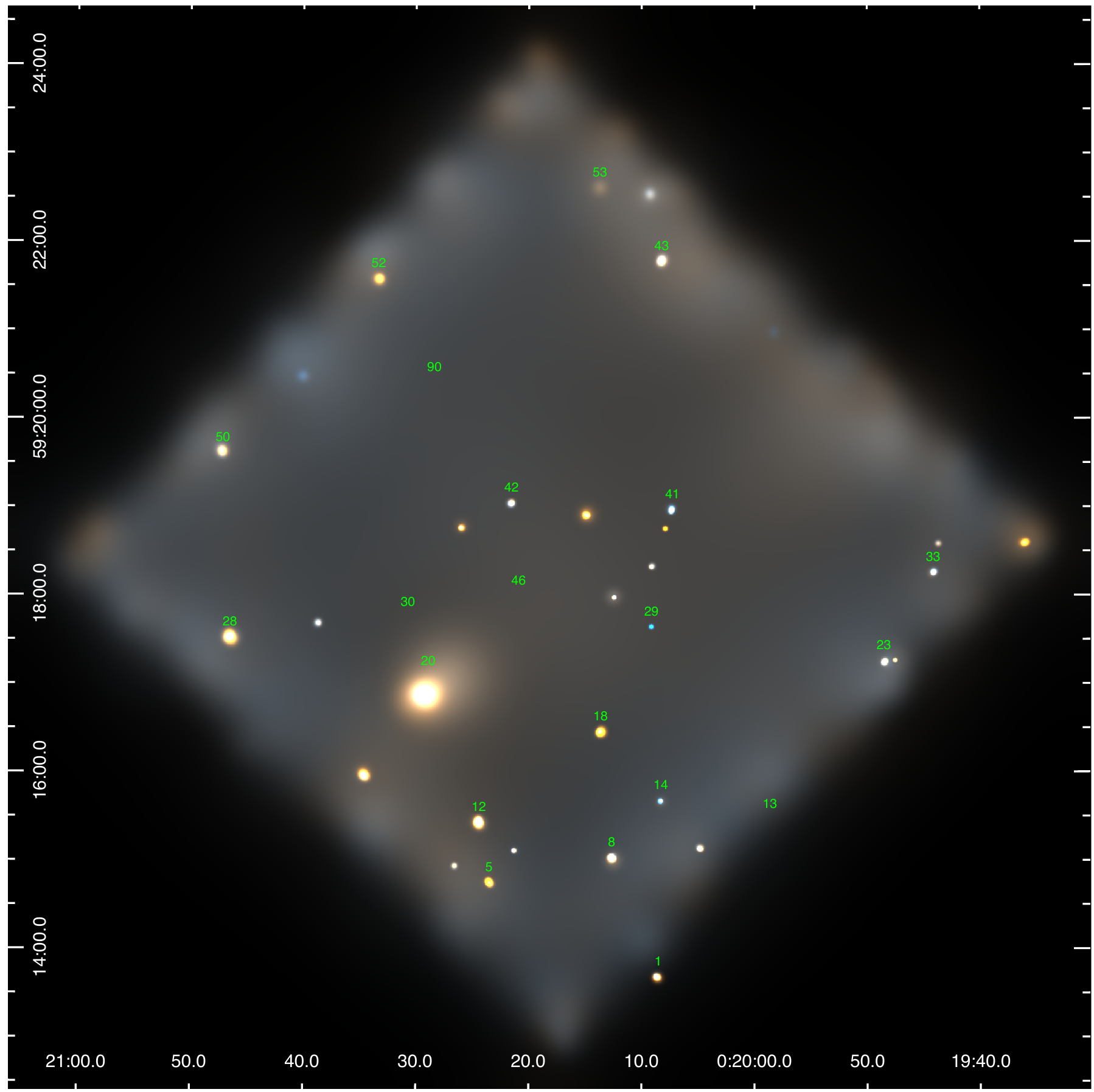}
\caption{ \chandra ~X-ray image of IC 10   ({\it obsid 57082}) with adaptive smoothing and representative RGB color-map. The RGB channels are red = 0.3-1.5 keV, green = 0.3-8 keV, blue= 2.5-8 keV, so that harder sources  will appear blueish (e.g. \#29), and soft sources yellow/orange (e.g \#5).  The variable point sources listed in Table~\ref{tab:variables} are indicated by numbers. Coordinates are RA, Dec in J2000 epoch. The field of view is the same as Figure~\ref{fig:cxo_images}.}
\label{fig:cxo_image}
\end{center}
\end{figure*}

Images for visual inspection were generated using both gaussian-smoothed and adaptively-smoothed  B (0.3-8 keV), S (0.3-1.5 keV), and H (2.5-8 keV) band data. Smoothing is necessary to create an image from sparse data. The adaptive smoothing maps created with {\it ciao smooth} for the B-band images were applied to make the S \& H band images, and all 3 bands used the exposure maps generated for 1.5 keV photons (roughly the peak of ACIS sensitivity). To illustrate our search for transient sources we present all 10 B-band images in Figure~\ref{fig:cxo_images} which shows a number of prominent transient events and also illustrates the effects of shifting spacecraft roll-angle and aim-point. The individual variable sources can be identified with the aid of Figure~\ref{fig:cxo_image} which is a representative color image constructed using the visualization and analysis program {\it SAOimage DS9}\footnote{http://ds9.si.edu} from the  deep merged dataset 57082 (Table~\ref{tab:dataset}).

\section{Constructing the Source Catalog} 
\label{sect:catalog}

For each \chandra ~observation we ran {\it Ciao} {\it wavdetect} on each of the three energy bands (S, B, H). {\it wavdetect} was run with the default significance threshold of $10^{-6}$ at wavelet scales of 1, 2, 4, 8, and 16 pixels. Each detection is given its own source $srcid$, and  the detections in each band are cross-matched by position (95\% uncertainty radius computed by the formula of \citealt{Hong2005}) to produce a list of unique sources within the observation,  linked by an identifier called $src\_uid$. The list derived from the 90 ksec stacked $obsid$ 57082 was bore-sighted against the optical catalog of \cite{Massey2007}. Source lists from each $obsid$ were then corrected to the same reference frame, using the coordinates of IC 10 X-1 (by far the brightest source). We then compiled the final point-source catalog from the entire set of observations by using the tool {\it classifytable} in the {\it starbase} relational database package\footnote{http://hopper.si.edu/wiki/mmti/Starbase}.  {\it classifytable} is used to group together sources that lie within a certain distance (5'') of each other, these matches are then filtered using the 95\% uncertainty radii. All detections belonging to a unique source are linked by the identifier {\it SrcNumber}. Coordinates and error radius are inherited from the $srcid$ with the ``best" detection (most counts, smallest error radius). The number of unique X-ray point sources detected in each observation is listed in the final column of Table~\ref{tab:dataset}.

Next, the unique source list for each observation was bore-sighted to the unique source list obtained from the deep merged dataset of ObsIDs 07082 \& 08485 (which had identical pointing). This reference data set had itself already been bore-sighted to the optical catalog of \cite{Massey2007}. The bore-sighting procedure between X-ray observations used only the brightest source (IC 10 X-1) due to its very small statistical positional uncertainty arising from its large count rate. 

Once bore-sighting was complete, we constructed a master catalog using a 5 arcsec matching radius around each source ($src\_uid$). Finally we applied a filter to eliminate false matches: requiring the matching $src\_uid$ coordinates to lie within the quadrature sum of their 95\% error radii and an additional 0.75" arcsec accounting for statistical and systematic errors. The resulting master catalog contains 110 unique point sources, many of which have multiple detections. All detections of the same source share the same {\it SrcNumber}  identifier and a \chandra~ source name  based on the usual convention (CXOU J HHMMSS$\pm$DDMMSS) taking the coordinates from the $srcid$ with the smallest error circle. 

The \chandra ~observations all have slightly different fields of view as a consequence of the operational requirements of the spacecraft. A certain pitch and roll angle are required to maintain the proper orientation with respect to the Sun. We worked with the mission planners to adjust the exact pointing for each observation once the pitch/roll combination was known. This enabled us to keep most of the same point sources within the ACIS-S3 CCD during each observation.  The master catalog was processed with a {\it Unix} script that called {\it Ciao} to query the exposure map for the coordinates of every source in each observation to determine whether the source was actually observed with ACIS-S3 or not. If the source was observed {\it and not detected} by {\it wavdetect}, then a new row was added to the master catalog with the Modified Julian Date (MJD) of the observation, the value $src\_significance$ = ``NULL" and an estimated upper-limit for the broad-band count rate. This upper limit was computed as the equivalent count-rate that would generate 5 net counts at the HRMA aim-point according to PIMMS 
(Portable Multi-Mission Simulator \footnote{http://heasarc.gsfc.nasa.gov/cgi$-$bin/Tools/w3pimms/w3pimms.pl}) for an absorbed power-law spectrum ($\Gamma=1.5, ~N_H = 5\times10^{21}$ \nh) divided by the normalized exposure-map value at the source location.  The simulations of ~\cite{Kim2004} indicate this is a $\sim$75\% detection probability for the off-axis distances encountered within the S3 chip, at low background rates. If the source was found to lie outside the S3 chip, then no modifications were made to the table. The master source catalog is maintained as a single table, combining all detections and upper limits, from which light curves are extracted as described in the next section. 

Positional matching to the catalog of \cite{Wang2005} was performed using our and their  95\% error radii in quadrature; the resulting IDs are provided in Table~\ref{tab:variables}. To facilitate observations at other wavelengths, and cross-correlation with other databases we provide our complete point-source catalog as supplemental data in the online journal.

 \subsection{Light-Curve Generation}
\label{sect:lc}

Light curves for all of the 110 unique sources were constructed by processing the master catalog with a {\it Unix} script which computed the following parameters for each group of rows bearing the same $SrcNumber$ identifier: Number of occasions observed ($N_P$), number of detections in the broad-band ($N_B$), minimum count rate ($R_B^{min}$), maximum count rate ($R_B^{max}$), average count rate ($R_B^{mean}$); root-mean-square $R_B^{RMS}$; variability range ($Range$ = $R_B^{max}$ - $R_B^{min}$); variability ratio ($Ratio = F_B^{max}/F_B^{min}$); 
relative variability ($\sigma_{Var} = Range/ \sqrt{(errorR_B^{max})^2 + (errorR_B^{min})^2}$); and total number of net broad-band counts ($Counts_B^{Total}$) obtained by summing the values from each detection.  In computing the above parameters we included all B-band detections and B-band upper limits, except that 
$F_B^{max}$ (and its associated error bar) are required for a positive detection. The complete set of parameters characterizing each source are provided in the online journal as supplemental data accompanying Table~\ref{tab:variables}, which shows only the variable objects, whose selection is described below. 

\begin{figure*}
\begin{center}
\includegraphics[angle=0,width=16cm]{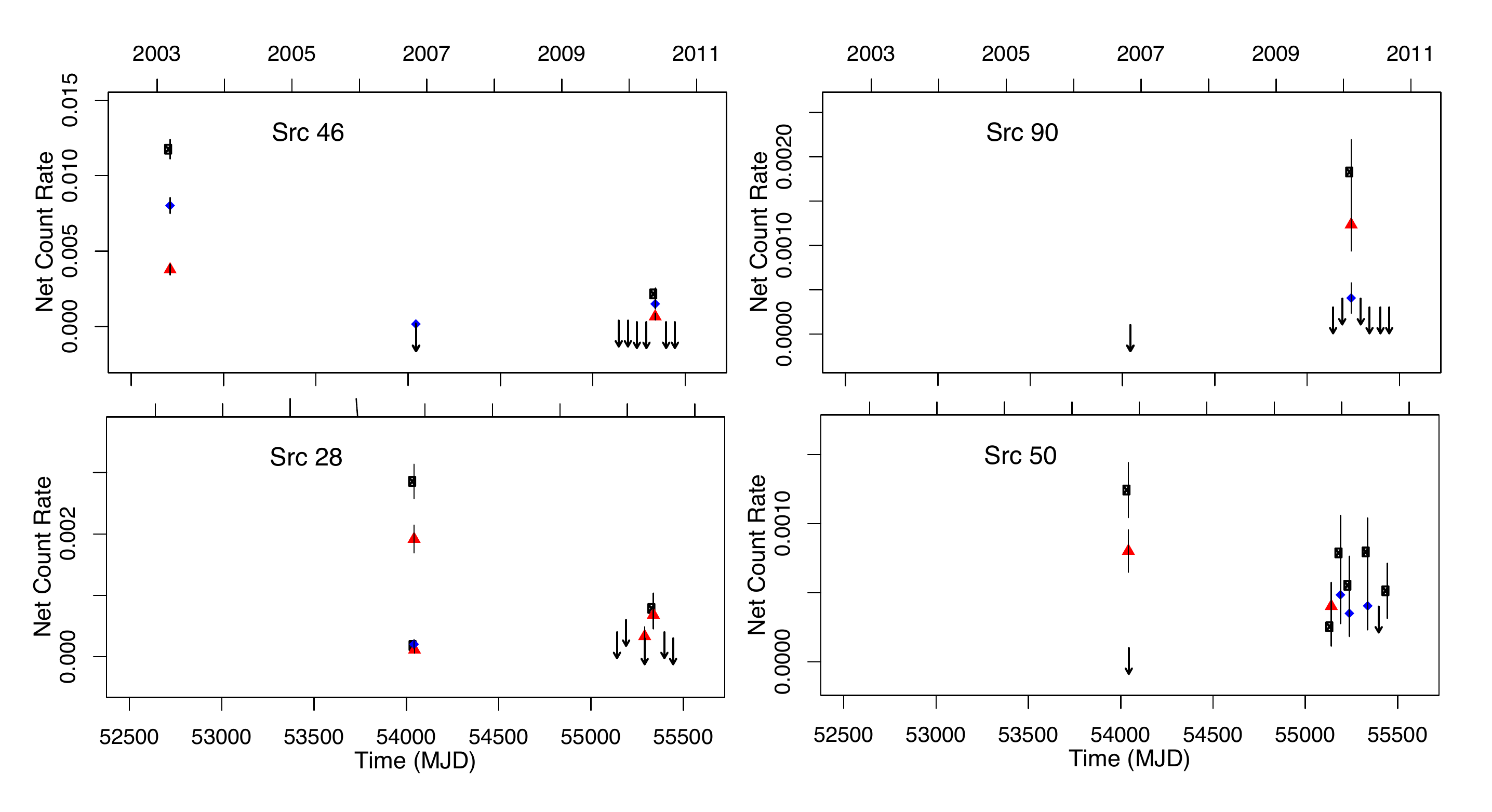}
\caption{ Light-curves of strongly variable X-ray sources with $Ratio>10$ and $\sigma_{Var}>3$, 2003-2010. Note that Src 46 is IC 10 X-2. Colors denote energy bands: broad B=0.3-8 kev (black), soft S=0.3-1.5 kev (red), hard H=2.5-8 kev (blue). Upper limits for B-band in the case of non-detection are indicated by arrows. Lightcurves drawn from the ACIS-S3 detector in the 10  (unstacked) $obsids$ listed in Table~\ref{tab:dataset}. Not all sources were in the FoV for all observations, hence the differing time coverage and number of points. Refer to Table~\ref{tab:variables} for a summary of the derived lightcurve parameters. Supplemental data-behind-the-figure for each light-curve are available in the online-journal.}
\label{fig:lcs_gt10}
\end{center}
\end{figure*}

\begin{figure*}
\begin{center}
\includegraphics[angle=0,width=16cm]{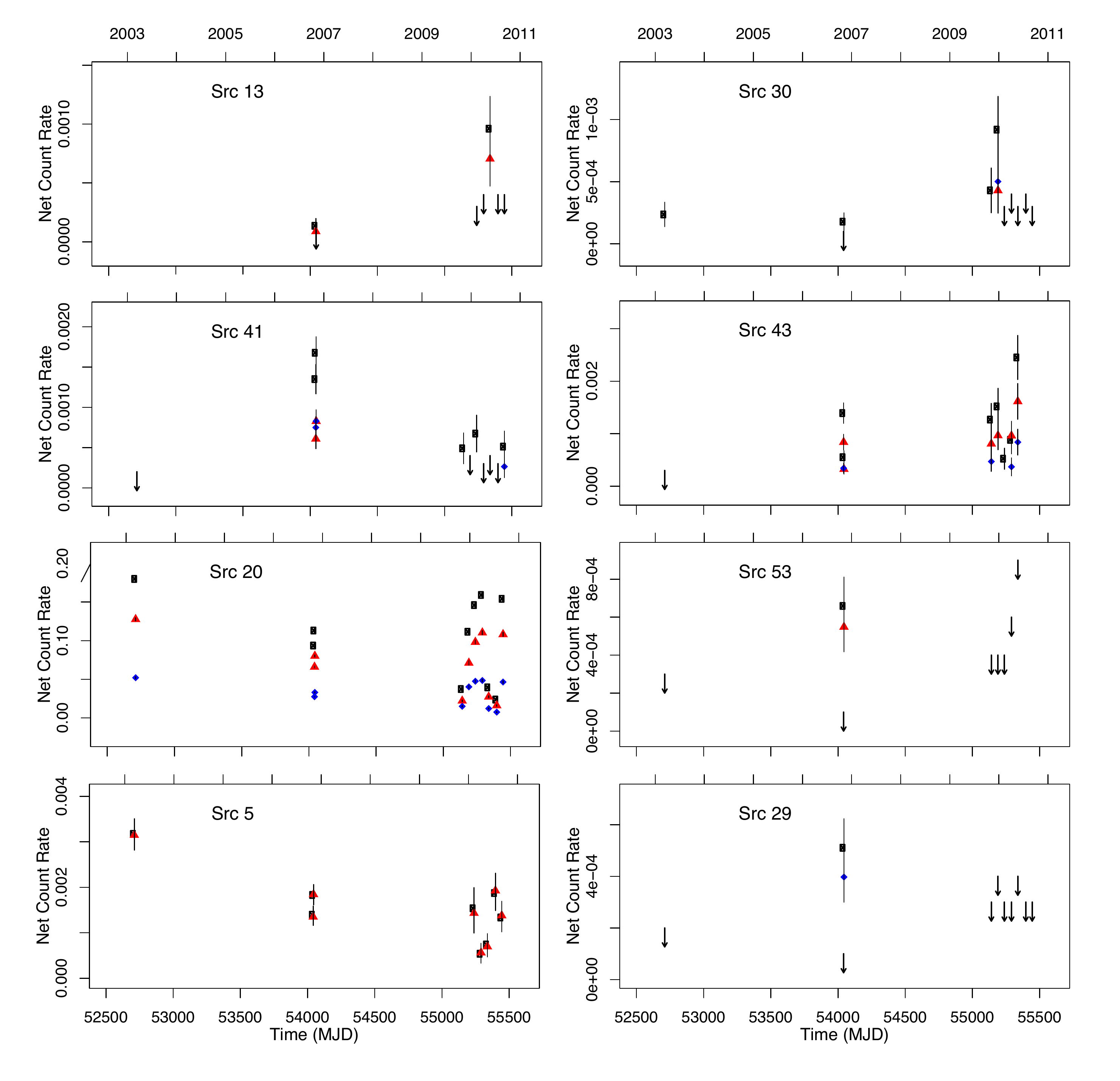}
\caption{Light curves of Moderately Variable X-ray Sources selected according to  $5<Ratio<10$ and $\sigma_{Var}>3$. Note that Src 20 is IC 10 X-1. Plot symbols denote B (black), S (red), H (blue) energy bands as Figure~\ref{fig:lcs_gt10}.
Supplemental data-behind-the-figure for each light-curve are available in the online-journal.}
\label{fig:lcs_gt5}
\end{center}
\end{figure*}

\begin{figure*}
\begin{center}
\includegraphics[angle=0,width=16cm]{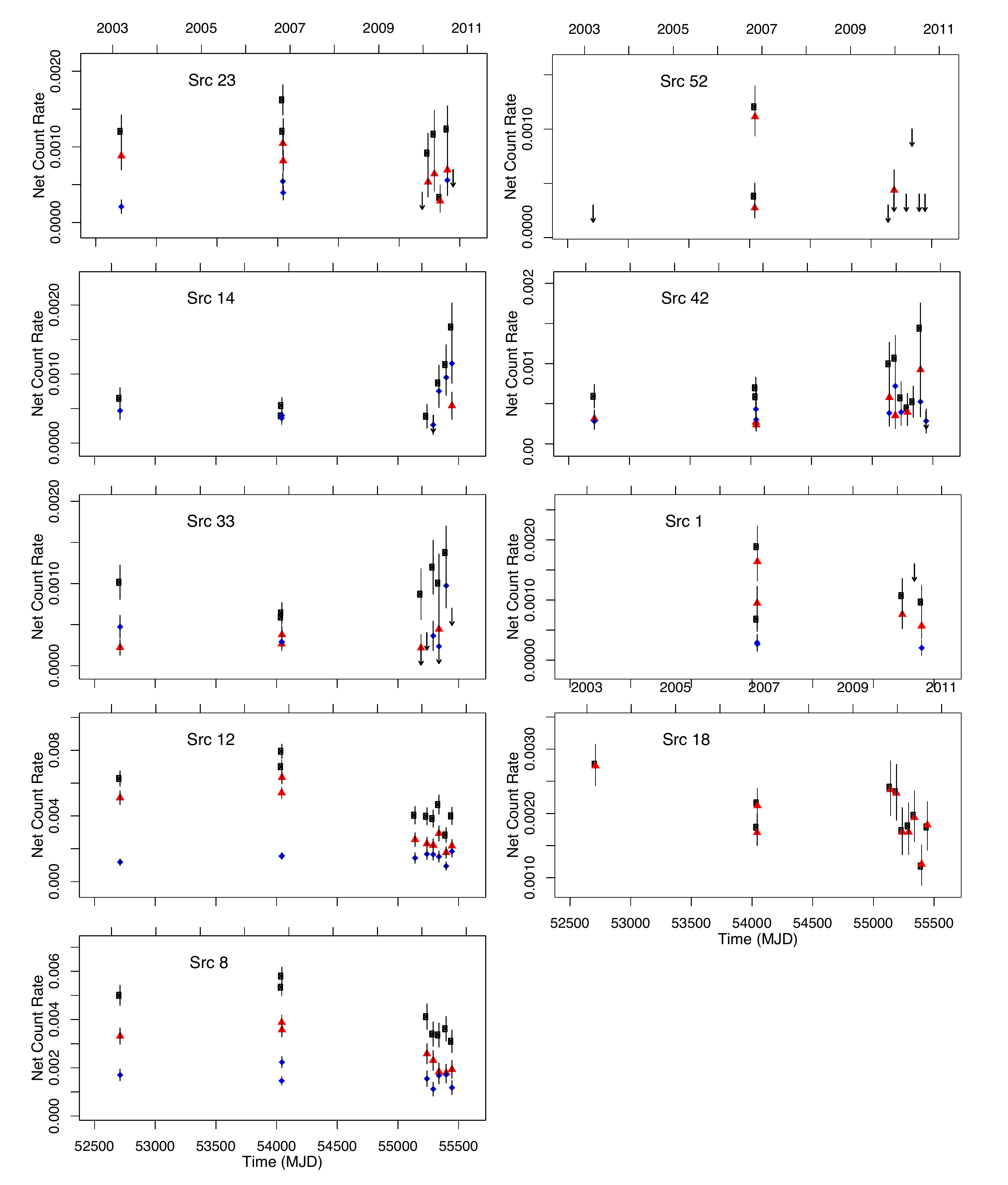}
\caption{Light curves of Variable X-ray Sources selected according to $Ratio<5$ and $\sigma_{Var}>3$. Plot symbols denote B (black), S (red), H (blue) energy bands as Figure~\ref{fig:lcs_gt10}. Supplemental data-behind-the-figure for each light-curve are available in the online-journal.}
\label{fig:lcs_lt5}
\end{center}
\end{figure*}

Light curves were plotted for all X-ray sources in the master catalog; selected examples are shown in Figures~\ref{fig:lcs_gt10},~\ref{fig:lcs_gt5}, and~\ref{fig:lcs_lt5}. The plots display detections in all 3 energy bands and upper limits for observations with no B-band detection. All 110 light curves were visually examined alongside the {\it ACIS} images to verify the detections and to note any abnormalities such as sources falling very close to the edge of a CCD or having a close neighbor. In this way, we flagged 8 sources as ``probably false" (objects detected only once at the extreme edge of the CCD) and 4 sources as being duplicates (multiple detections of a real source being falsely identified as separate sources when detected in subsequent observations). These eventualities are expected, and the number of cases was within the expected range.  {\it wavdetect} expects to generate one false source per CCD (i.e. $\sim$ 10 in total) and faint off-axis sources are subject to larger positional errors. We note that most of the sources were detected in at least two independent observations. Specifically, 43 sources were detected more than once in the B-band, 36 only once in B-band (some of which were also detected in a separate observation in S or H only), and 25 were only ever detected in the S or H band.  The probability of a source with multiple detections being false is extremely small given the low surface density of sources.

The light-curve parameters described above (and given in Table~\ref{tab:variables}) were used to identify variable and transient sources by sorting on the $\sigma_{Var}$ and $Ratio$ parameters. Assuming the usual statistical definitions, a light curve with $\sigma_{Var} > 3$ is variable at the 99\% confidence level. There are 21 such objects, listed in Table ~\ref{tab:variables} ranked in descending order by $\sigma_{Var} > 3$.   Our criterion for detecting a variable or transient source is at least one B-band detection with count rate at least $3\sigma$ above the lowest upper limit; thus, only real sources are included in Table ~\ref{tab:variables}.

Complete data for each variable source are provided in tabular form as ``data-behind-the figure'' accompanying this paper in the online journal.

\begin{deluxetable*}{lllllllllllll}
\tablecaption{Chandra ACIS-S Source Catalog of IC 10. 
 \label{tab:variables}} 
\tablehead{ \colhead{Source}  & \colhead{W05} &   \colhead{$N_P$}     & \colhead{$N_{B}$}  & \colhead{$\bar{R_B}$}           & \colhead{$R_B^{Min}$}                        &   \colhead{$R_B^{Max}$}                    & \colhead{Range}                  & \colhead{Ratio} & \colhead{$\sigma_{var}$} & \colhead{Total Net Cts} & \colhead{$E50$}  & \colhead{Class}\\
                                        &                &               &                                       &   \colhead{$ct~ks^{-1}$}   &   \colhead{$ct~ks^{-1}$}  &   \colhead{$ct~ks^{-1}$}  &   \colhead{$ct~ks^{-1}$}   & 			 &                                &  & \colhead{keV} &  \\   }
\startdata

46   & 11C &  10  &  2    &  1.61    &  0.1     &  11.75    &  11.65    &  117.5  &  18.2  &  370  & 3.35 & T, X\\
90  & - &  9  &  1  &  0.45  &  0.1  &  1.83  &  1.73  &  18.3  &  4.7  &  26  & 1.64 &  T, S \\
28  & 24C &  8  &  3  &  0.73  &  0.18  &  2.86  &  2.67  &  15.6  &  9.2  &  130  & 1.13 & T, S \\
50  & 17C, 46X & 8  &  6  &  0.58  &  0.1  &  1.24  &  1.14  &  12.4  &  5.7  &  88 & 2.07 &  T, X?\\
13  & - & 7  &  2  &  0.39  &  0.1  &  0.96  &  0.86  &  9.6  &  3.1  &  19  & 1.69  & TC,S? \\
30  & 19C & 10  &  4  &  0.36  &  0.1  &  0.92  &  0.82  &  9.2  &  3.1  &  33  & 1.97 & TC, X?\\
41  & - & 10  &  5  &  0.63  &  0.2  &  1.68  &  1.48  &  8.4  &  7.4  &  154 & 2.53 & TC, X\\
43  & 24X & 8  &  7  &  1.11  &  0.3  &  2.45  &  2.15  &  8.2  &  5.1  &  170  & 2.06  & V, X\\
20   & 18C, 39X &10  &  10  &  105.8  &  23.7  &  179.87  &  156.17  &  7.6      &  55.6  &  22632 & 1.75 & V, X \\
53  & - & 8  &  1  &  0.47  &  0.1  &  0.66  &  0.56  &  6.6  &  3.7  &  27  & 1.38 & TC, U \\
5  & 14C, 35X & 8  &  8  &  1.55  &  0.54  &  3.17  &  2.63  &  5.9  &  6.7  &  292  & 0.89 & V, U\\
29  & 4C & 10  &  1  &  0.31  &  0.1  &  0.51  &  0.41  &  5.1  &  3.6  &  22  & 3.60 & TC, X\\
23  & 17X & 9  &  7  &  0.98  &  0.33  &  1.62  &  1.29  &  4.9  &  4.9  &  203  & 1.86  & V, U\\
52  & - & 9  &  2  &  0.53  &  0.3  &  1.21  &  0.91  &  4.0  &  4.7  &  64  & 1.24 & V, S\\
14  & 3C, 25X & 8  &  7  &  0.76  &  0.39  &  1.68  &  1.29  &  4.3  &  3.3  &  117  & 3.64 &V, X \\
42  & 12C, 33X & 10  &  9  &  0.73  &  0.4  &  1.44  &  1.04  &  3.6  &  3.3  &  144  & 2.70 & V, X/A?\\
33  & 16X & 9  &  7  &  0.87  &  0.4  &  1.38  &  0.98  &  3.4  &  3.0  &  132  & 2.55  & V, X/A?\\
1  & 3C, 36X & 5  &  4  &  1.24  &  0.68  &  1.89  &  1.21  &  2.8  &  3.0  &  70  & 1.76 & V, X\\
12  & 15C, 36X & 9  &  9  &  4.96  &  2.84  &  7.93  &  5.1  &  2.8  &  8.1  &  1137  & 1.69  & V, U\\
18  & 8C & 10  &  10  &  1.99  &  1.18  &  2.77  &  1.59  &  2.3  &  3.6  &  435  & 1.12  & V, U\\
8  & 7C, 28X & 8  &  8  &  4.22  &  3.1  &  5.8  &  2.7  &  1.9  &  4.5  &  863  & 1.83 &  V, X\\

\enddata
\tablecomments {Objects in this table were selected as variable sources accord to $\sigma_{Var}>3$  and are sorted by $Ratio$ in descending order. $N_P$ = number of observations where the source fell on the ACIS-S3 detector; $N_B$ =  number of broad-band detections; the remaining columns are computed from those broad-band detections and the upper limits derived for the non-detections. $Class$ is a candidate source classification based on the X-ray and optical properties: T = Transient ($Ratio >$ 10), TC = Transient Candidate,  V = Variable, X = X-ray binary, S = Stellar Corona, A=AGN, U = unclassified.  Column $W05$ is the source identifier from \cite{Wang2005} where the suffix $C$ and $X$ denote \chandra ~and \xmm ~respectively.  
{\it A full version of this table containing all 110 unique sources, their celestial coordinates, and error bars for all columns is provided in the online journal.} }
\end{deluxetable*}

\subsection{X-ray Spectra and Energy Quantiles}
\label{sect:quantile}

\begin{figure}
\includegraphics[angle=0,width=8.5cm,trim={0cm 0.5cm 0 0cm}]{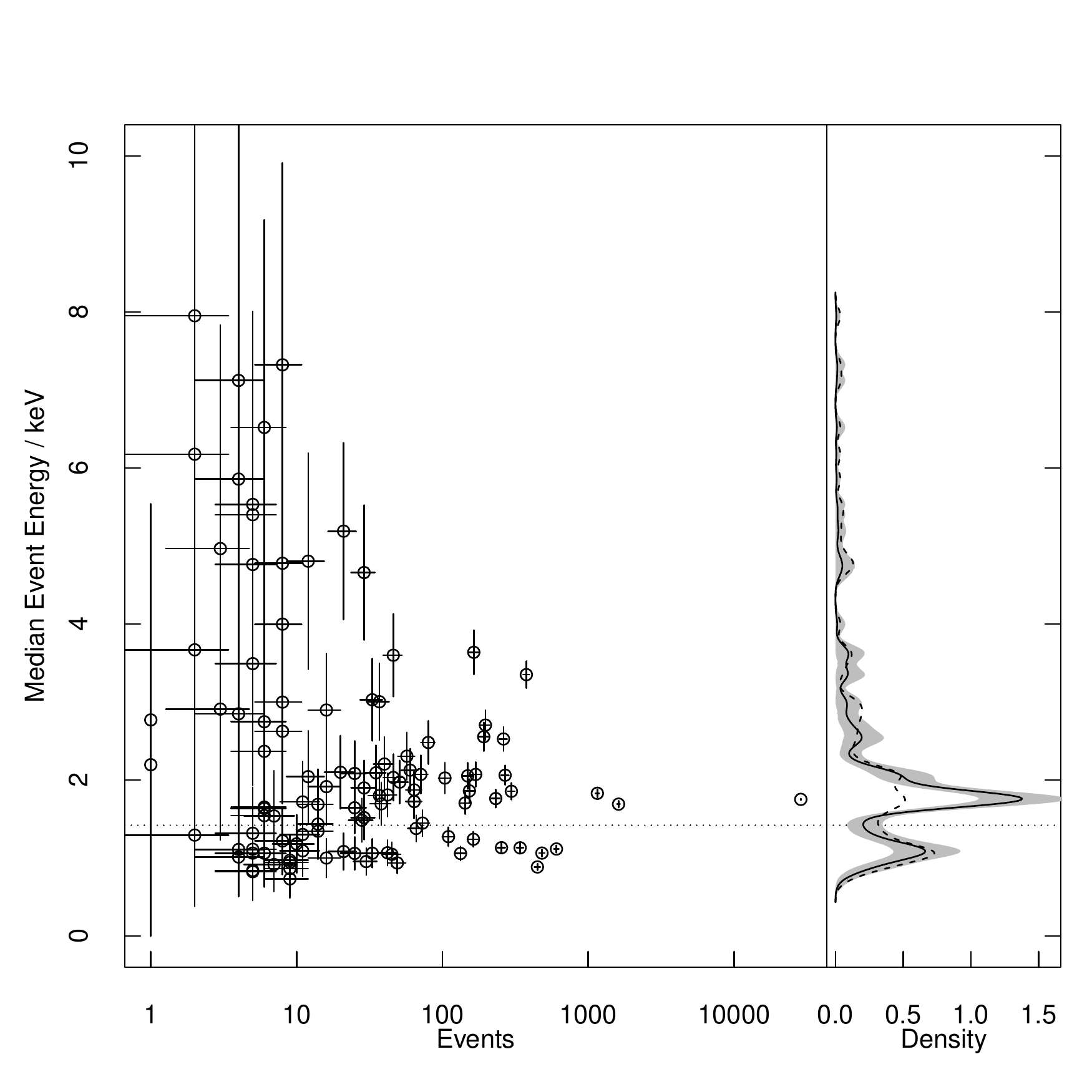}
\caption{Hardness-Intensity Diagram for all sources in the ACIS-S3 catalog. Filled points are 3$\sigma$ variable sources identified by inter-observation photometry provided in Table~\ref{tab:variables}, those in red have large amplitudes ($Ratio > 5$). The distribution of objects as a function of median event energy is projected in the right-hand panel using a Kernel Density Estimator (bandwidth 0.1 keV) with and without error weighting (solid and dashed lines respectively). The 95\% confidence region (grey) was obtained by bootstrap sampling. The distribution is bimodal with a split at around 1.4 keV. We suggest that objects in the soft peak are likely to be foreground objects, and those in the hard peak IC 10 members, due to the significant absorbing column in front of IC 10. }
\label{fig:hardint}
\end{figure}

Given the large distance to IC 10 we are able to extract spectra for relatively few sources. Consequently the median energy for the combined events is reported in Table~\ref{tab:variables} as an indication of spectral hardness. We plotted a hardness intensity diagram for all 110 point sources in Figure~\ref{fig:hardint}, using cumulative events as a proxy for intensity, and the median event energy for the hardness axis. This parameter space serves to illustrate some key properties of the objects in our catalog. Figure~\ref{fig:hardint} shows that large amplitude variability is seen at all intensity levels, that most  bright sources are moderately variable ($\sigma_{var} >3$), and that below 100 counts our selection method (Section~\ref{sect:lc}) is sensitive only to large amplitude variables. Thus we are not picking up random fluctuations in faint sources as variables. Figure~\ref{fig:hardint} also shows that the distribution of  median event energy is bimodal, with a hard/soft split at 1.4 keV. This was revealed by a kernel density estimator ({\it density} in the {\it R} package, with bandwidth = 0.1 keV) which we ran both with and without weighting by the errors. Weighting increased the strength of the bimodality, as seen by comparing the solid and dashed lines in the right-hand panel of Figure 6. We performed a bootstrap resampling (10$^4$ trials) using the error-weighted values to generate 95\% confidence intervals for the density plot. The {\it kmeans} clustering test identifies a similar hard-soft split, which can be attributed to different extinction values for IC 10 vs foreground galactic objects. 

Energy quantiles \citep{Hong2004} were computed from the event lists for each unique source in the master catalog. Quantiles were computed for each observation and also for the combined event list, which provides the maximum number of counts and hence delivers better S/N values. These values and the accumulated number of events in the source extraction regions is listed in the supplementary table in the online journal. In our analysis we use two of the quantiles defined by \citep{Hong2004}: the Slope quantile $QDx$ which is related to the steepness of the spectrum, and the Curvature quantile $QDy$ which is constructed from the ratio of the quartiles $Q_{25}/Q_{75}$. Full details are available in \cite{Hong2004} and in the source code.\footnote{https://hea-www.harvard.edu/ChaMPlane/quantile/}.

\begin{figure*}
\begin{center}
\includegraphics[angle=0,width=14cm]{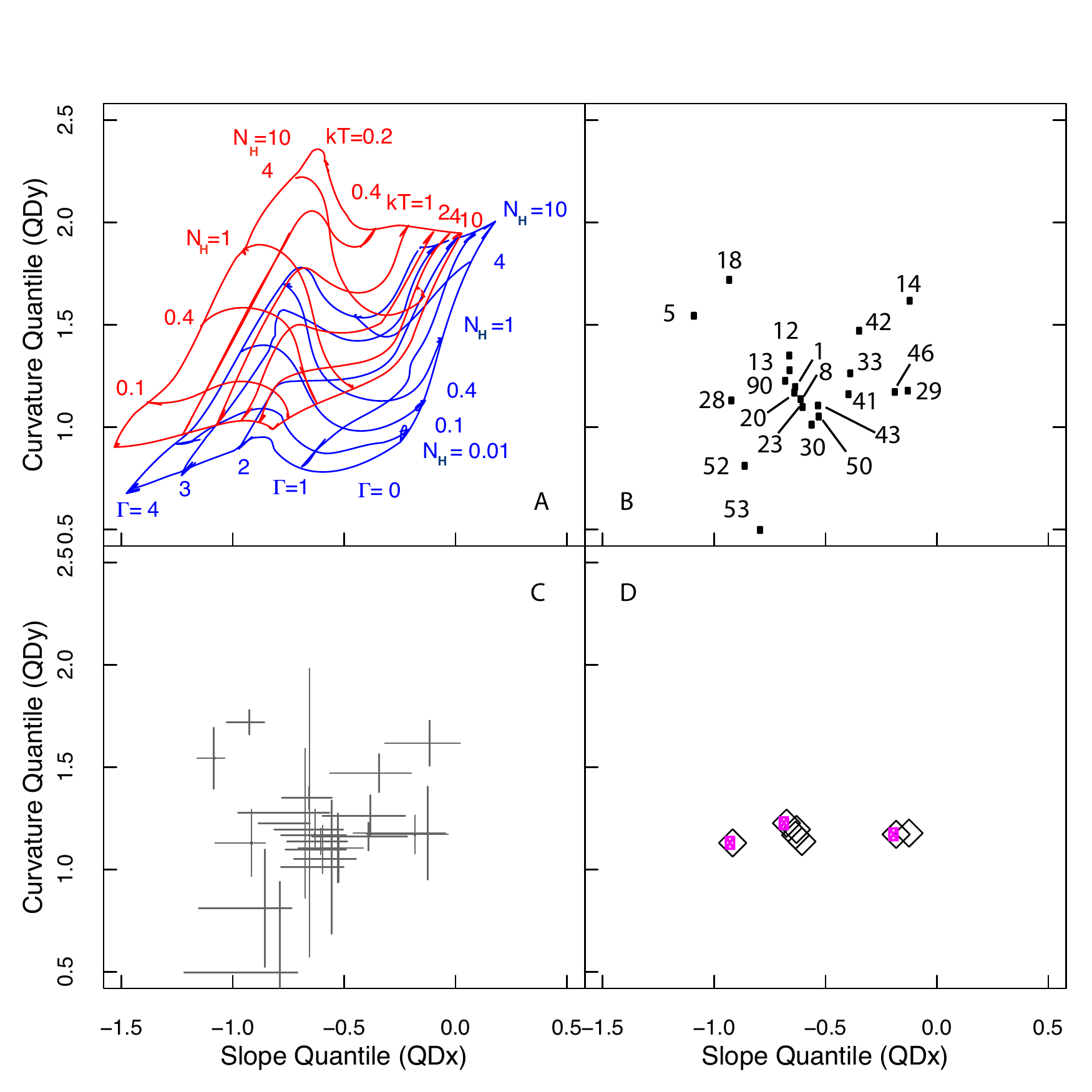}
\caption{X-ray Energy Quantiles for all variable point sources in IC 10 (Table~\ref{tab:variables}), generated from the merged event lists of all 10 observations. All panels share the same scale which indicates spectral hardness in the X-axis and curvature in the Y-axis. Increasing absorption causes a spectrum to turn over at low energies, thus increasing curvature is associated with rising $N_H$. The panels are as follows: (A) Model Grids, (B) Quantile values, (C) error bars, and (D) objects with optical counterparts in Table~\ref{tab:optical}, with H$\alpha$ emitters shown in magenta color. The majority of sources cluster in the X-ray Binary region characterized by absorbed power-law models with $\Gamma \simeq 1-2 and N_H \simeq (0.5-1) \times 10^{22}$ \nh. Sources toward the left likely have thermal spectra, while sources on the upper right (none of which have optical counterparts) could be absorbed XRBs or background AGN. }
\label{fig:quantiles}
\end{center}
\end{figure*}

The $QDx$, $QDy$ parameter space shown in Figure~\ref{fig:quantiles} was designed to explore in a single analysis, the spectral shape of low count sources that differ widely in hardness. Panel A shows spectral model grids generated for the back-illuminated {\it ACIS-S3} detector for two classes of sources: power law (photon index $\Gamma= 0-4$) and thermal bremsstrahlung ($kT=0.2-10 keV$), both of which with absorption varying between $N_H = (0.01-10)\times10^{22}$\nh. These models are intended to represent accretion-powered and stellar coronal sources, respectively. The quantile values for the variable sources selected in Table~\ref{tab:variables} are plotted in panel B, with labels to facilitate discussion. Error bars are in panel C, and those sources with optical counterparts (table~\ref{tab:optical}) are highlighted in panel D.  

\begin{figure*}
\begin{center}
\includegraphics[angle=0,width=16cm]{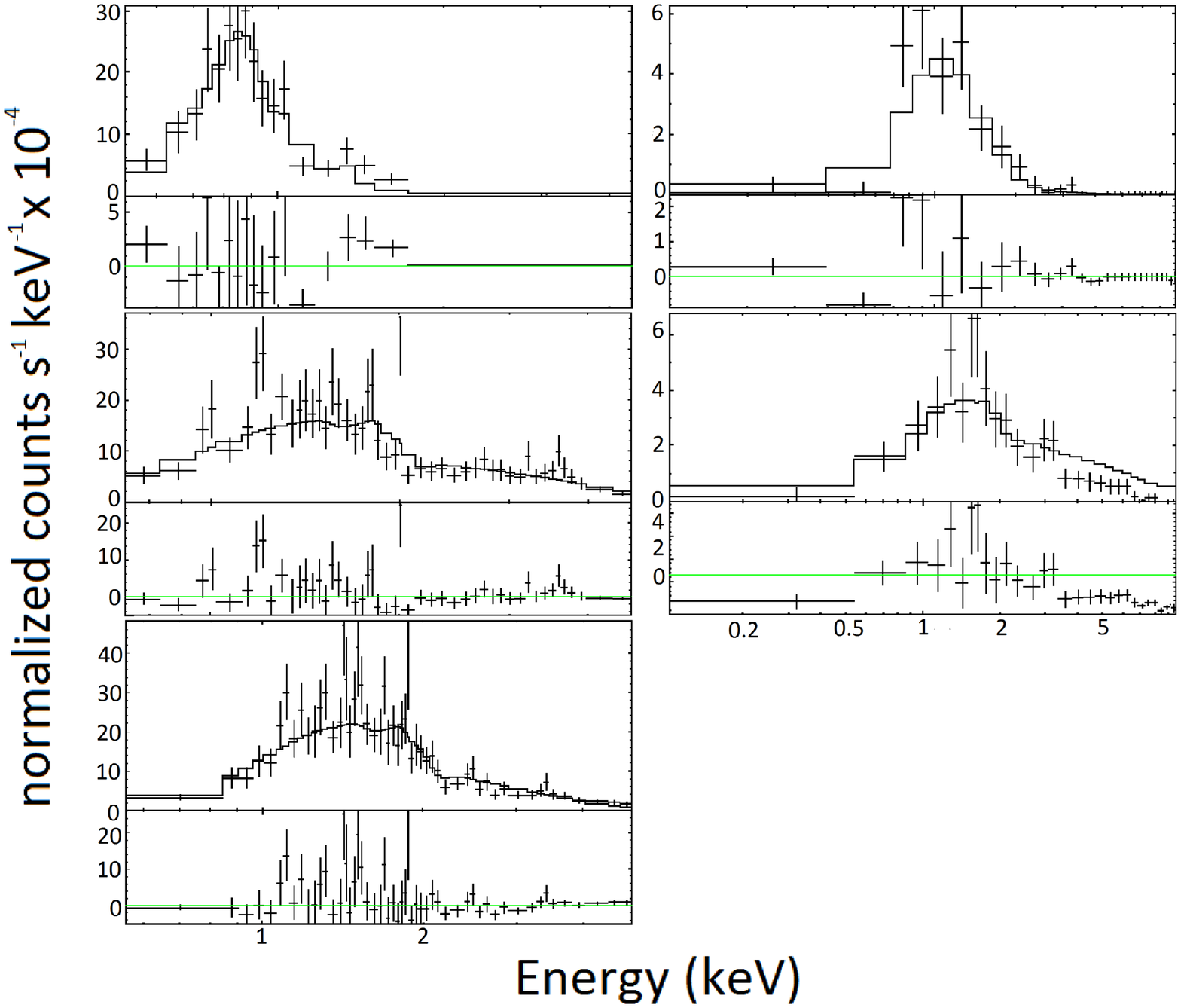} 

\caption{CCD X-ray spectra of sources 5, 8, 12, 18, and 23 (clockwise from top left) are shown with their best-fitting single-component models and residuals.  All spectra are plotted for the energy range 0.2-8 keV, with flux and residuals per spectral bin in units of $counts s^{-1} keV^{-1}$ 
Data from multiple observations were merged for this analysis, with the models  folded through the combined response files generated by {\it Ciao Specextract}. See Table~\ref{tab:xrayspec} for details of the models and their parameters.}
\label{fig:spectra}
\end{center}
\end{figure*}

Pulse-height spectra were extracted for all sources having greater than 200 net counts in the master catalog, yielding 7 objects (there are 13 sources with $>$ 100 net counts). In this article, we include only those appearing in Table~\ref{tab:variables} after excluding IC 10 X-1 and X-2 (which have been reported elsewhere), which leaves 5 objects for the analysis in this paper. For every such object, we used the {\it Ciao} script {\it specextract} to generate source and background spectra and response files (RMF, ARF) for each observation; and also for the merged event list containing all observations in which the source was detected.   The resulting spectra are plotted in Figure~\ref{fig:spectra}. The spectra were examined in XSPEC\footnote{https://heasarc.gsfc.nasa.gov/xanadu/xspec/} and were fitted by the following models (combined with the {\it phabs} absorption model): power-law (PL), Raymond-Smith (RS), Mekal (M), black-body (BB), and thermal bremmstrahlung (TB). These models were selected as appropriate for X-ray spectra of accretion, magnetic flare, and thermal origin.  In cases where the absorption was unconstrained by the model we froze it at 5$\times10^{21}$ \nh ~for the column density toward IC 10.  The resulting model parameters and goodness of fit statistics are presented in Table~\ref{tab:xrayspec}. Given the low S/N ratios, only single component fits were considered.  We caution that both quantiles (in common with all flux ratio-based color systems) and single-component spectral fits can be non-unique. For example spuriously low $N_H$ values can occur when soft components remain unaccounted for \citep{Brassington2010}. This is a motivation to obtain deeper datasets where such components can be resolved.


\begin{deluxetable*}{cccccc}
\tablecolumns{8}
\tablecaption{X-ray Spectral Model Fits
 \label{tab:xrayspec}} 
\tablehead{ \colhead{Model} &  \colhead{$N_H\times 10^{22}$} & \colhead{$kT$ } & \colhead{$\Gamma$} & \colhead{Norm$\times 10^{-5}$} & \colhead{$\chi^{2}_{red}/$DoF} \\ 
                   \colhead{}           & \colhead{\nh}                              & \colhead{keV} & \colhead{} & \colhead{} & \colhead{} \\ }
 \startdata
Source 5 \\
PL  & 0.842 $\pm$ 0.3 & - & 7.5 $\pm$ 2.2 & 2.0 $\pm$ 0.9 & 1.167/74  \\
RS & 0.442 $\pm$ 0.1 & 0.28 $\pm$ 0.06 & - & 2.1 $\pm$ 1.2 & 0.596/74  \\
M & 0.000 $\pm$ 0.1 & 0.55 $\pm$ 0.08 & - & 0.21 $\pm$ 0.09 & 0.631/74  \\
BB & 0.006 $\pm$ 0.1 & 2.8 $\pm$ 0.02 & - & 7.6 $\pm$ 0.8 & 1.829/74  \\
TB & 0.78 $\pm$ 0.5 & 0.13 $\pm$ 0.04 & - & 0.02 $\pm$ 0.001 & 0.934/74  \\
\tableline
Source 8 \\
PL & 0.565 $\pm$ 0.09 & - & 1.6 $\pm$ 0.2 & 1.3 $\pm$ 0.2 & 1.020/53  \\
RS & 0.483 $\pm$ 0.07 & 10.2 $\pm$ 4.2 & - & 4.2 $\pm$ 0.3 & 1.011/53  \\
M & 0.482 $\pm$ 0.07 & 10.4 $\pm$ 4.4 & - & 4.2 $\pm$ 0.2 & 1.017/53  \\
BB & 0.051 $\pm$ 0.06 & 0.92 $\pm$ 0.005 & - & 0.061 $\pm$ 0.004 & 1.049/53  \\
TB & 0.49 $\pm$ 0.07 & 9.4 $\pm$ 3.8 & - & 1.5 $\pm$ 0.1 & 1.001/53  \\
\tableline
Source 12 \\
PL& 0.73 $\pm$ 0.08 & - & 2.2 $\pm$ 0.1 & 2.9 $\pm$ 0.4 & 0.835/64  \\
RS & 0.57 $\pm$ 0.06 & 4.1 $\pm$ 0.5 & - & 5.8 $\pm$ 0.4 & 0.866/64  \\
M & 0.57 $\pm$ 0.06 & 4.1 $\pm$ 0.6 & - & 5.7 $\pm$ 0.4 & 0.883/64  \\
BB & 0.19 $\pm$ 0.05 & 0.71 $\pm$ 0.03 & - & 0.065 $\pm$ 0.3 & 1.180/64  \\
TB & 0.58 $\pm$ 0.05 & 3.6 $\pm$ 0.7 & - & 2.5 $\pm$ 0.3 & 0.849/64  \\
\tableline
Source 18\\
PL & 0.831 $\pm$ 0.2 & - & 6.2 $\pm$ 2.1 & 0.41 $\pm$ 0.2 & 0.993/46  \\
RS & 0.072 $\pm$ 0.03 & 6.2 $\pm$ 1.8 & - & 0.16 $\pm$ 0.09 & 1.013/46  \\
M & 0.623 $\pm$ 0.2 & 0.56 $\pm$ 0.2 & - & 0.19 $\pm$ 0.1 & 0.990/46 \\
BB & 0.000 $\pm$ 0.04 & 0.32 $\pm$ 0.05 & - & 0.001 $\pm$ 0.0004 & 1.004/46  \\
TB & 0.453 $\pm$ 0.1 & 0.39 $\pm$ 0.1 & - & 1.17 $\pm$ 0.9 & 0.9934/46  \\
\tableline
Source 23 \\
PL & 2.45 $\pm$ 0.9 & - & 5.8 $\pm$ 0.6 & 5.4 $\pm$ 1.2 & 0.803/71  \\
RS & 3.31 $\pm$ 0.5 & 0.21 $\pm$ 0.06 & - & 0.002 $\pm$ 0.0005 & 0.805/71  \\
M & 0.834 $\pm$ 0.3 & 1.06 $\pm$ 0.9 & - & 0.95 $\pm$ 0.6 & 0.834/71 \\
BB & 0.001 $\pm$ 0.002 & 3.31 $\pm$ 0.9 & - & 0.15 $\pm$ 0.08 & 0.827/71  \\
TB & 1.02 $\pm$ 0.8 & 1.7 $\pm$ 1.2 & - & 0.42 $\pm$ 0.2 & 0.810/71  \\

\enddata
\tablecomments{Spectral fits for each variable source with $>$200 counts. Models are a simple multiplicative combination of {\it phabs} and one of: PL=Powerlaw, RS=Raymond-Smith, M= Mekal, BB= Blackbody, TB=Thermal Bremsstrahlung}
\end{deluxetable*}


\subsection{Optical Counterparts}
\label{sect:optical}
A list of positional counterparts from \cite{Massey2007} was generated by the procedure described in \cite{Laycock2014}.   A limiting visual magnitude of $V>23.8$ was used for sources without counterparts, this being the faintest magnitude of any object in the catalog.  The distance independent X-ray to optical flux ratios were found using

\begin{equation}
\label{eqn:fxfv}
     log_{10}\frac{f_x}{f_v} =  log_{10}(f_x)+ \frac{V}{2.5} + 5.37\ ,             
\end{equation}
where the observed flux $f_x$ was computed from the net count rate assuming an absorbed power-law with a photon index of 1.5 and $N_H=5 \times10^{21}$ \nh. The values in Table~\ref{tab:optical} are the maximum measured values, remembering that $f_x$ is variable.

In the following analysis, sources with optical counterparts believed to lie in IC 10 have their apparent magnitudes corrected for the foreground extinction assuming a value of $A_v$= 2.85 and distance modulus $\mu=24$. 
Narrow-band $H\alpha$ and  $H\alpha~continuum$ images from the Gemini Observatory were examined in order to look for sources of Balmer-line emission as described in \cite{Laycock2014}. Preliminary spectroscopic followup was reported by \cite{Balchunas2011}, although spectral type and radial velocity (RV) are only available for 3 of the variable sources.

\section{Discussion of Individual Variable X-ray Sources}
\label{sect:discussion}
\subsection{Transients}
Four sources have variability $Ratio>10$ and $\sigma_{var}>3$, these are Src 46 (IC 10 X-2), Src 90, Src 28, \& Src 50 (listed in order of decreasing amplitude). The light curves for these sources appear in Figure~\ref{fig:lcs_gt10}.  There are a further 2 sources (Src 53 and Src 29) that we include as transients as each was detected in a single observation, with a broad-band count rate exceeding the lowest upper limit by three $\sigma$ (i.e. $N_B=1$ and $\sigma_{Var}<3$, but not reaching $Ratio>10$) listed in Table~\ref{tab:variables}. Both were in the ACIS-S3 field of view 8-10 times, so that the null detections and corresponding upper limits place meaningful limits on the duty-cycle. Lightcurves for these two objects are include in Figure~\ref{fig:lcs_gt5}. Only one transient, Src 46,  which is IC 10 X-2, \cite{Laycock2014} provided enough counts to produce pulse-height spectra and appears to be an X-ray binary with a supergiant Be companion and similarities to Supergiant Fast X-ray Transients (SFXTs). For reference we include its lightcurve in Figure~\ref{fig:lcs_gt10} and it appears on the quantile diagram (Figure~\ref{fig:quantiles}) at (-0.18,1.17) which is consistent with a very hard power-law spectrum. The case of IC 10 X-2 highlights the need for optical spectra, since its photometric colors are unremarkable unless one knows the appropriate correction, and its nature as an extragalactic HMXB required the unambiguous doppler-shift to place it firmly in IC 10.   

\subsection{Src 28. [W05]24C}
Src 28 was observed 8 times yielding 3 B-band (and one S-band only) detections with a variability ratio of 15.6, a 9.3$\sigma$ variation above the most constraining upper limit. Quantile values for the 4 observations were all consistent with the merged value. Lying at [-0.9,1.1] in Figure~\ref{fig:quantiles} the source is consistent with a soft power law ($\gamma \sim2.5$) or a $\sim$few keV thermal plasma. In both cases the absorption is small ($<<10^{21}$\nh) suggesting a foreground galactic location for the object (subject to the caveats in Section~\ref{sect:quantile}.) There are two optical counterparts (Table~\ref{tab:optical} from \citealt{Massey2007}) and the source is not in the field of the Gemini images, so no followup spectrum was obtained. Only one of these stars is consistent with the X-ray detection having the smallest error circle. Both stars have color indices suggestive of spectral type M. Taking the most likely counterpart, whose B-V is already at the un-reddened value for M0V, we calculate distance modulus $\mu \simeq 18.79  - 9 = 9.8$. The distance is then $\sim$900 pc, yielding peak $L_x = 3.7\times10^{30}$\lx, in the range of isolated dMe stars, which is supported by the distance independent ratio \fxfv =-0.53 being lower than seen for active binaries. Note that in the absence of a spectral type or RV value, IC 10 membership cannot be entirely ruled out. If located at 660 kpc, and de-reddened by $A_V$=2.85 the star would have an extreme absolute magnitude $M_V\gtrsim-8$, which favors the galactic dMe interpretation.

\subsection{Src 29, [W05]4C}
Src 29 was detected in {\it obsid 08458} with 22 net counts which placed it a factor of 5 (or 3.6$\sigma$) above its most sensitive non-detection. The detection is in the B and H (and not S) bands. The quantile value [-0.13,1.2] lies in the hard power law region ($\Gamma\sim0.5$)  of Figure~\ref{fig:quantiles} on the contour for $N_H \sim 10^{22}$ \nh.  There are 2 optical counterparts in M07 having near identical photometry ($V =$ 21.7, $B-V \simeq +0.93$), Gemini $H\alpha$ line and continuum images indicate that two stars are blended in the \cite{Massey2007} images, with one catalog position on top of the brighter star, and the other midway between the two. The source lies in a region containing clumpy diffuse $H\alpha$ emission, but neither star is a strong $H\alpha$ point-source. A GMOS spectrum of the brighter star (which is near the center of the X-ray error circle) shows a strong continuum without obviously identifiable features other than telluric lines. This rules out a foreground M-dwarf and suggests a reddened early-type star in IC 10. Applying the distance modulus and extinction correction assumed throughout this paper gives $M_V = 21.7 -2.85 - 24 = -5$, and $(B-V)_o \simeq 0$. The star is also blue in U-B and V-R color indices compared to the other stars in Table~\ref{tab:optical}, and once de-reddened looks like a B spectral type. This object is likely an HMXB in IC 10, based on X-ray and optical properties.

\subsection{Src 50 [W05]17C,46X}
Src 50 was observed 8 times, yielding 6 B-band detections and two upper limits. The most constraining upper limit comes from one of the pair of 45 ksec exposures made in Nov 2006 providing the evidence for a large variability ratio ($ratio$=12, $\sigma_{Var}$=5.7). The quantile values varied with time,  the spectrum being harder in {\it obsids 11081,11082,11084}. The merged value plotted in Figure~\ref{fig:quantiles} lies at [-0.52,1.05] which is dominated by photons from its softer apparition in 2006. The model grid indicates a power law ($\Gamma \sim 1$) spectrum with typical IC 10 absorption ($\sim$ 5$\times10^{21}$\nh). There is an optical counterpart in \cite{Massey2007} but it lies outside the Gemini field and no followup spectrum has been obtained. If the object is in IC 10, then its de-reddened photometry gives $M_V=22.5 - 2.8 - 24.01 = -4.3$, $B-V=1.823 - 0.86 = 0.9$. 

\subsection{Src 53}
Src 53 was detected only in {\it obsid 08458} with 27 net counts, at a count-rate a factor of 6.6 ( 3.7$\sigma$ ) above the most constraining non-detection {\it obsid 07082} which was of equal exposure-time (45 ksec) and pointing,  and occurred only a day before.  A point-source is clearly visible in the image of {\it obsid 08458} and absent in {\it obsid 07082}. The detection was made in the S and B bands only. The very rapid variability and soft spectrum implied by the S-band detection are indicative of a magnetic stellar flare but not conclusive. The quantile value [-0.8,0.5] lies outside our model grids and has large errors so we can say little about its spectral shape.  There is an optical counterpart (V=21.4, Table~\ref{tab:optical}) but it lies outside the Gemini field and no followup spectrum has yet been obtained. 

 \subsection{Src 90}
Src 90 was covered by the ACIS-S field of view 9 times, but was detected only once, in {\it obsid 11082} in the S, B \& H bands. It is clearly visible as a point source in the adaptively smoothed image. The observed count-rate was a factor of 18 (4.7$\sigma$) above the most constraining upper limit. The energy quantile (Figure~\ref{fig:quantiles}) lies at [-0.67,1.23] within the power law model grid at a location consistent with $\gamma \sim 2$, $N_H\sim 10^{21}$ \nh.
There is an optical counterpart (\ref{tab:optical}) which we identify as an $H\alpha$ emission line star in Gemini narrow-band imaging. We obtained a Gemini GMOS optical spectrum of this star \cite{Balchunas2011}, which shows a red continuum with molecular bands characteristic of spectral type M. Prominent lines of $H\alpha$,  $H\beta$ in emission enable us to measure the radial velocity  RV=-165 km/s, RV=-171 km/s for the respective lines. Thus the star belongs to the Milky Way since the mean RV of IC10 is -340 km/s.  Photometry from \cite{Massey2007} (V=21.348, B-V=1.8) indicates the color-index is consistent with early M. For example spectral type M0Ve ($B-V=+1.5, M_V=9)$  would require only E(B-V)$\sim$0.3, $A_V\lesssim 1$. The distance modulus is then $\mu \simeq 21.4 - 1 - 9 = 11.4$, for a distance of $\sim$2 kpc, implying $L_X \simeq 1.2\times10^{31}$\lx. The distance-independent flux ratio during the flare reaches \fxfv = +0.3. Stellar flares in dMe stars reach $L_X \simeq 10^{31}$, with the BY Dra binary type being particularly luminous \cite{Singh1996}. The luminosity, quantile, and duty-cycle point to a hard and energetic event on a magnetically active M-dwarf. Such interlopers are frequent bi-catch of X-ray surveys, and we have seen one in the foreground of the SMC \cite{Laycock2009}.

\subsection{Persistent Variables}
Many of the brighter sources in our study show significant variability ($\sigma_{Var} > 3$)  as expected given that astrophysical X-ray emission mechanisms are dynamic in nature. The light curves of these candidates are presented in Figure~\ref{fig:lcs_gt5} for those varying by a factor of 5-10, and in Figure~\ref{fig:lcs_lt5} for those not exceeding a factor of 5.  All of these objects have sufficient counts in the merged event-list to provide energy quantiles with reasonable S/N.  In Figure~\ref{fig:quantiles} we plot the quantiles for all  unique objects in the master catalog having at least 10 counts in their merged event list, with the candidate variable sources picked out in larger points.  A group of 10 candidate variables clusters along the $N_H$ contour corresponding to $\sim5\times10^{21}$\nh within the power law model grid region indices in the range $\Gamma=1-2$.  Based on this evidence they are consistent with being HMXBs in IC10.   In the following section we present individual analyses for the remaining variability candidates listed in Table~\ref{tab:variables}, beginning with the five candidate variables to exceed 200 net counts. For these objects we performed spectral fitting in XSPEC testing PL, BB, TB, NEI models  combined with the photo-absorption model {\it phabs}.  The best fit for each of these sources is displayed in table \ref{tab:xrayspec}.

\subsection{Src 1 = [W05]3C,36X}
Source 1 was observed in half of the ACIS-S observations and is detected in 4 of the 5. The are enough counts for the quantile diagram to place it firmly in the XRB locus. The optical counterpart $V=22.4, B-V=0.7$ if adjusted for distance ($\mu=24$) and reddening ($A_V=2.85$) gives $M_V=-4.5, (B-V)_0=-0.16$ which implies an early B spectral type. Thus the object is an HMXB, subject to optical spectral confirmation of the precise subtype.

\subsection{Src 5 = [W05]14C,45X }
The light-curve for Src 5 (figure~\ref{fig:lcs_gt5}) shows a large amplitude variable with a 100\% duty cycle, being detected in all observations for which it was in the FoV. It also shows that virtually all the counts are in the soft band. The hardness-intensity plot (Figure~\ref{fig:hardint}) shows Src 5 is at once the softest variable, and one of the brightest objects in the survey, pointing to a foreground object.  Quantile analysis suggests a thermal spectrum with median energy of just 0.5 keV that did not vary between detections.  The $\sim300$ net-count spectrum was best fit by a black-body model (Fig. \ref{fig:spectra}, Table~\ref{tab:xrayspec}).  A power-law model can provide an acceptable fit but the spectral index of 7.7 implies the cool thermal plasma is more physically motivated. No optical counterpart is found in \cite{Massey2007} and the source lies outside the Gemini field. 

\subsection{Src 8 = [W05]7C,28X }
Src 8 is a very low amplitude variable with 100\% duty cycle in the Chandra survey. The summed 863 net-count spectrum was fit with a power-law model of $\Gamma = 1.6$, and typical IC 10 absorption (Fig. \ref{fig:spectra}, Table~\ref{tab:xrayspec}).  There is an optical counterpart in \cite{Massey2007} (lies outside the Gemini field) for which we can estimate $M_V = 22.974 - 2.85 - 24  = -3.9$,  $(B-V)_o = 1.242 -  0.86 =   0.32$ which would fit a supergiant or Be-X interpretation.

\subsection{Src 12 = [W05]15C, 36X }
Src 12 is a persistent source with 100\% duty cycle and a lightcurve that suggests very slow X-ray variability. The source has the second highest number of total X-ray counts ($>$1000) in the survey, so spectral fitting was feasible. The quantiles lie at the upper end of the cluster of points in Figure~\ref{fig:quantiles}, and accordingly a powerlaw fit to the spectrum yields $\Gamma \simeq 2$ powerlaw with typical absorption for the IC 10 field (Figure~\ref{fig:spectra}, Table~\ref{tab:xrayspec}). Turning to physically motivated models, acceptable fits are obtained for the thermal plasma models (TB, RS, Mekal), all giving temperatures consistent with $kT=4 keV$, and $N_H =6\times10^{21}$\nh ~similar to the powerlaw fit. A blackbody fit yields a significantly lower column density and temperature (kT$\lesssim$1 keV). The lack of an optical counterpart in M07 provides the following constraints on distance-independent flux ratio (\fxfv~ $>1$.9) and absolute magnitude ($M_V>-3$) if located in IC 10. The object could be an AGN on the basis of the spectral index, large \fxfv~ and slow variability. A foreground dMe star seems unlikely due to the lack of flares and persistently high \fxfv. An XRB cannot be ruled out. Given the persistent nature of this source it will be possible to obtain much higher S/N X-ray spectra and timing in the future.

\subsection{Src 13}
A transient X-ray source appearing to exhibit rapid variability, Src 13 was detected (at a very low flux) in only one of the deep (45 ksec) observations separated by 2 days in 2006. Subsequently the source flared up in a single one of the 5 monitoring observations of 2009/10 which are spaced approx. 6 week apart. These two independent positive detections confirm the reality of the object. The limiting fluxes for those four nulls are at or above the flux detected in the deep 2006 exposure. Both positive detections were dominated by soft-band counts, with no detection in the hard band. The quantile value is consistent with a plasma temperature of $kT\simeq 1-2 keV$, or a $\Gamma\simeq2$ powerlaw. There is an optical counterpart within the 95\% confidence radius. The photometry (Table~\ref{tab:optical}: $V=20.5, B-V=1.5$) would be consistent with an un-reddened M0V star at a distance of 2 kpc, or a hotter spectral type at larger distance. It is impossible to make the star closer since the observed color index imposes the limit of $M_V\leq9$.  If located in IC 10 we obtain  $M_V$=-6.4. An optical spectrum is needed to resolve the nature of this source. 

\subsection{Src 14 = [W05]3C,25X }
Src 14 has an interesting lightcurve that contains an outburst of duration $\sim$2 months during 2010. Quantile analysis reveals a persistently hard source, having the highest median photon energy (E50 = 3.64 keV) of any object in Table~\ref{tab:variables}, and lying in the high-absorption hard power-law region of Figure~\ref{fig:quantiles}.  The lack of an optical counterpart down to a limiting magnitude of $\gtrsim$23.8 provides us with a distance independent flux ratio  \fxfv $ >$ +1.24 for the brightest recorded X-ray flux.
If located in IC10, the absolute magnitude would be $M_V = 23.8 - 2.85 - 24  = -3$ if we assume the typical extinction correction. A more luminous star would require more extinction to remain hidden.  Spectral type B0V has $M_V=-3$ so the data support a Be-XRB.  If a dMe star located in the Milky Way, then the spectral type would have to be late M (for M0V the luminosity distance implied by $V>23.8$ is greater than 9 kpc) and the X-ray quantiles are too persistently hard. In summary the long duration outburst, persistently hard spectrum, and high \fxfv~ point to an XRB.

\subsection{Src 18  = [W05]8C}
Src 18 is a low amplitude (less than a factor of 2) variable with 100\% duty cycle in the Chandra survey.  The total 435 net-count spectrum is soft, equivalent to powerlaw $\Gamma>5$ or cool thermal plasma for any reasonable $N_H$ (Fig. \ref{fig:spectra}, Table~\ref{tab:xrayspec}). There is no optical counterpart in M07, providing a lower limit on \fxfv~ $>$+1.12. The X-ray position lies at the center of a faint compact $H\alpha$ shell identified in narrow-band Gemini images. A GMOS slit placed across the ring did not attain sufficient S/N to detect any lines.

\subsection{Src 23 = [W05]17X }
Src 23 is a variable source with a high duty cycle (7/10) and no optical counterpart in M07 which puts a lower limit on \fxfv~ $>$+1.2.  With 203 net counts summed over the 6 observations where it was detected we obtained an acceptable X-ray spectral fit to the thermal bremsstrahlung spectral model with $kT$=1.7 keV, $N_H = 10^{22}$ \nh~ (Fig. \ref{fig:spectra}, Table~\ref{tab:xrayspec}).  Due to the low number of counts all simple models obtained similar $\chi^2$ values, with the powerlaw model confirming a soft spectrum ($\Gamma>5$) and the blackbody fit unable to constrain the $N_H$ to a significantly non-zero value. Indicative of a foreground galactic location.  

\subsection{Src 30 = [W05]19C}
Transient source with duty cycle $<50$ and no optical counterpart, if located in IC 10 the limiting magnitude $V>23.5$ implies \fxfv $>1, ~ M_V \simeq -3$ subject to a typical $N_H$. The quantile value is also consistent with an HMXB.

\subsection{Src 33 = [W05]16X}
Another high duty-cycle, hard, variable source without optical counterpart. As with Src 42, the object could be an HMXB or AGN with its high \fxfv $>$1.2, and limiting $V>23.5$ implying $M_V \simeq -3$ depending on absorption. 

\subsection{Src 41}
Hard transient source with duty cycle $<50$ and no optical counterpart, the lightcurve is reminiscent of a Be-XRB.  if located in IC 10 the limiting magnitude $V>23.5$ implies \fxfv $>1.23, ~M_V \simeq -3$ subject to a typical $N_H$. The quantile value is also consistent with an HMXB.

\subsection{Src 42 = [W05]12C, 33X}
This is a persistent variable source with a hard X-ray spectrum (median energy = 2.7 keV), and no optical counterpart in M07. Table~\ref{tab:variables} reports 9 broad-band detections out of 10 observations, however the X-ray lightcurve (Fig~\ref{fig:lcs_lt5}) shows a hard-band only detection in that observation (final point in 2010), so the source has a 100\% duty cycle. There are not enough counts to analyze the spectrum, however the quantile diagram Fig~\ref{fig:quantiles} shows Src 42 (the point at $QDx\sim-0.4, QDy\sim1.5$) is consistent with a heavily absorbed powerlaw $\Gamma \sim 2, N_H \sim 10^{22}$\nh. The limiting magnitude of $V\sim23.4$ constrains the optical counterpart to $M_V<-3$ for typical IC10 absorption, however the likely much higher $N_H$ could hide an O star. The flux ratio \fxfv  $>$+1.2 together with the hard spectrum and variability make an HMXB the most likely interpretation, although it could conceivably be a background AGN. 

\subsection{Src 43 = [W05]24X}
Variable source on all timescales probed, and seen in all observations except for the 2003 exposure (which places a limit on its quiescent flux), although since it appears as [W05]24X we know the source was seen by {\it XMM-Newton} in 2003. The lack of an optical counterpart limiting magnitude $V>23.5$ implies \fxfv $>1.4, ~M_V \simeq -3$ if located in IC10. The energy quantile is hard and lies in the HMXB region.

\subsection{Src 52}
This object is a large amplitude variable with 3 detections out of 9 observations. Two of these detections are mere days apart, and show a factor of 5 change in flux (perhaps a flare), with the fainter point only seen due to the long observation (45 ksec), while a 3rd and final detection was seen in the soft-band (only) in a single one of the 2009/10 monitoring series. There is a bright optical counterpart in M07, which at $V$=17.3, $B-V$=+1.6 is almost certainly a galactic star (the alternative being an extreme luminosity $M_V$=-9.6 hyper-giant if located in IC 10). Assuming spectral type M0 (based on the color index) the distance modulus is $\mu \simeq 17.3 - 9 \simeq 8.3$ giving a distance of 450 pc. The distance independent flux ratio \fxfv = -1.5 supports this interpretation but an optical spectrum is needed to confirm it.

\begin{deluxetable*}{lllllllllll}
\tablecaption{Optical Counterparts
 \label{tab:optical}} 
\tablehead{ \colhead{Src} & \colhead{Massey} & \colhead{$S$}    & \colhead{$r_{95}$} & \colhead{V}       & \colhead{B-V} & \colhead{U-B} & \colhead{V-R} & \colhead{R-I}     & \colhead{\fxfv} & \colhead{$H\alpha$} \\
                                         &                             & \colhead{arcsec} & \colhead{$arcsec$} & \colhead{mag} & \colhead{mag} & \colhead{mag} & \colhead{mag} & \colhead{mag} &                                 & \colhead{}     \\ }
\startdata
1   & J002008.68+591340.7 & 0.15 & 0.43 & 22.384 & 0.700 & -0.087    & 0.762 & 0.855  & 0.77    & n \\
8   & J002012.68+591501.4 & 0.1   & 0.32 & 22.974 & 1.242 &               & 0.735 & 1.144  &  1.45  & n\\
13 & J001958.69+591527.7 & 0.37 & 0.53 & 20.473 & 1.525 &               & 0.932 & 0.962  &  -0.33   &   n\\
20 & J002029.08+591651.7 & 0.30 & 0.28 & 21.722 & 0.905 & -1.094   & 0.137 & 0.754   &  2.43    & \\
20 & J002029.12+591651.8 & 0.17 & 0.28 & 22.478 & 0.017 & -0.916   & 0.851 & 0.805   &  2.74    & y \\ 
28 & J002046.55+591731.7 & 0.61 & 0.53 & 18.793 & 1.521 & 1.327    & 0.957 & 0.958   &   -0.53   & \\
28 & J002046.36+591732.4 & 0.55 & 0.80 & 20.199 & 1.786 &  99.999 & 1.191 &  1.395  &  -0.53    & n \\
29 & J002009.14+591738.3 & 0.32 & 0.39 & 21.777 & 1.059 & -0.820   & 0.536 & 1.052   &   -0.09  & n \\
29 & J002009.15+591738.4 & 0.37 & 0.39 & 21.732 & 0.932 & -0.588   & 0.494 & 1.049   &  -0.1     &  n \\
46 & J002020.94+591759.3 & 0.28 & 0.31 & 19.954 & 1.211 & -0.259   & 1.043 & 0.820   &  0.54    &  y\\
50 & J002047.28+591935.7 & 0.89 & 0.66 & 22.539 & 1.823 &              & 1.292 & 1.564   &  0.6      & \\
52 & J002033.29+592135.4 & 1.16 & 0.75 & 17.277 & 1.551 & 1.365   & 0.946 &              &  -1.5     & \\
53 & J002013.70+592236.1 & 0.49 & 0.96 & 21.404 & 1.872 &             & 1.354 & 1.603    &   -0.2    & \\
90 & J002028.42+592024.0 & 0.34 & 0.54 & 21.348 & 1.805 &             & 1.294 & 1.579    & 0.37     & y  \\
18 & *                                    &         &          &           &            &             &           &              &            & y \\


\enddata
\tablecomments{Positional optical counterparts from the catalog of \cite{Massey2007} for variable X-ray sources listed in Table~\ref{tab:variables}.  Matching criterion was Separation $S < \sqrt{ r_{95}^2 + 1}$.  $H\alpha$ column refers to the presence or absence of $H\alpha$ emission in Gemini GMOS narrow-band images which did not cover all sources.  Column \fxfv was computed from column $V$ and $R_B^{max}$ (Table 2) via eq.~\ref{eqn:fxfv}. *object 18 is coincident with ring-like $H\alpha$ structure.   }
\end{deluxetable*}

\section{Conclusions}
\label{sect:conclusions}

From a total of 110 unique point sources detected in 10 \chandra ~observations (years 2003-2010), we found 21 that are variable at the 3$\sigma$ level. The X-ray light curves exhibit a range of qualitatively different behaviors, including transients, large amplitude variables, and persistent sources with duty cycles ranging up to 100\%.  Analysis in the X-ray quantile diagram reveals a split between soft thermal sources and harder power-law like spectra that cluster along the model grid contour consistent with the average column density towards IC 10. While this separation is not a complete diagnostic of source classes, it supports to a large extent the hypothesis of \cite{Wang2005} that IC 10 hosts numerous HMXBs. Our variability study, better positions, and the increased spectral S/N ratios made possible by analyzing a cumulative 225 ksec of Chandra exposure (vs the previous 30 ksec) have provided a richer and more detailed view of this population. In this paper, we were able to confidently classify most of the variable sources and to usefully constrain the nature of the remaining sources. Deeper observations are needed to increase the number of individual sources for which X-ray spectral fits can be obtained.  Variability is a characteristic of all known astrophysical X-ray production mechanisms, and the fact that the majority of the catalog sources do not make the  3$\sigma$ cut is due to their faintness (and consequent difficulty in establishing variability) of most objects encountered.

The HMXB candidates presented in this paper are all characterized by power-law spectra, strong X-ray variability, and either the presence of an optical counterpart consistent with an O/B star in IC10; or the lack of an optical counterpart to a limiting magnitude of $V\gtrsim23.8$ which still allows for the existence of an early type star (B0V or fainter, $M_V>-3$ for minimal IC 10 extinction $A_V=2.85$). Substantially higher extinction values do exist in parts of IC 10 as there is a high degree of structure in the interstellar medium (ISM) of that galaxy.   
In the former category, we find Src 1, Src 8, and Src 29, in addition to the well-studied Src 20 =  IC 10 X-1 and Src 46 = IC 10 X-2, for a total of 5 HMXBs with optical counterparts, only one of which has a known orbital period and three have no optical spectrum.

Of the sample in this paper, only 12 X-ray sources have positional optical counterparts in \cite{Massey2007}, among which we identify 4 sources with $H\alpha$ line emission. Optical spectra have been obtained for just 4 counterparts to X-ray variables. These were the known BH+WR binary IC 10 X-1; the previously reported HMXB IC 10 X-2; a reddened early-type star associated with the hard absorbed transient source \#29; and a foreground dMe star associated with the soft transient source \#90.  

Given that highly luminous stars exist in IC 10, spectroscopic followup is required to positively identify the majority of the optical counterparts to these sources. The photometric colors of the counterparts are expected to be somewhat degenerate between extinction and spectral type due to the large line-of-sight depth through the foreground galactic plane.   Future efforts must be directed at obtaining optical spectra of the counterparts which can be efficiently done with multi-object spectrographs on 8-m telescopes thanks to the compact size of IC10. Positive spectroscopic identification of the X-ray binaries requires spectral types, confirmed IC 10 membership via radial velocity, and ultimately orbital elements from RV and/or photometry. The distance and extinction values to IC 10 indicate that only the upper-end of the spectral sequence can be resolved from ground-based telescopes, however this is not an insurmountable problem since HMXBs are formed from precisely these stars.  This paper reports objects with X-ray properties similar to known HMXBs, both with and without candidate optical counterparts. No transients were seen at super-Eddington luminosity, thus no candidate BH X-ray novae occurred during any of the observations. 

Transient X-ray sources are conventionally identified by a variability ratio in the $>$10 range and transient candidates in the 5-10 range (e.g. \cite{Williams2008}). We are aware of powerful and innovative approaches that rely on Bayesian analysis (e.g. \citealt{Brassington2012}) to obtain the probability that the variability ratio for an individual source lies within the range of interest. These techniques reduce bias and hence offer two advantages: (a) enhanced sensitivity to extremely faint transients even in the case of a very small number of observations, (b) reduced tendency to falsely identify positive photon-counting fluctuations as variability.  Our selection criteria are based on the conventional statistical prescription for comparing data points. In particular the 3$\sigma$ cut is conservative (99\% significance) and avoids sweeping up low probability transient candidates. With 110 point sources detected, the expectation is of order 1 false variable candidate out of the 21 variables identified.  Admittedly, the cost of this conservative approach is reduced sensitivity to fainter transients. There are 36 single-detection B-band sources, only 2 of which made it into the sample presented here.  As we showed in Figure~\ref{fig:hardint} all of those 34 other objects are very faint, therefore they are not HMXBs in outburst. 

As to the nature of the new HMXBs reported here, we can already assert that they are a different population than that in the Magellanic clouds. The young age of the underlying stellar population and the elevated production of massive stars and stellar remnants noted by other authors (e.g. \cite{Massey2002, BB}) have implications for the expected X-ray binary population. The SMC with its episodic starburst history contains at least 100 known or suspected HMXBs, the vast majority of which are Be+NS systems, and all earlier than B3 \citep{McBride2008}. It was first proposed by \cite{Negueruela1996} that the narrow spectral type range seen in Be-HMXB is an evolutionary signature. This idea is supported by the groundbreaking work of \cite{Antoniou2010}, who showed the SMC Be-HMXB are associated with distinct stellar populations of age 40-70 Myr and \cite{Williams2013} who found a similar age-association for HMXBs in NGC 300 and NGC 2403. In IC 10 there has not been sufficient time to form these objects, therefore we expect to see completely different HMXB species. Following this logic, we assume that the Be phenomenon is not yet widespread in IC 10, hence accretion-powered X-rays must involve mass donors with stronger winds and/or smaller orbital separation in order to deliver the requisite mass-transfer rate. Supergiants and WR stars are the obvious candidates, although we find only one X-ray source (X-1) coincident with the WR catalog of \cite{Crowther2003}; weak-lined WR stars could be present but as-yet unidentified.  Some very young HMXBs have recently been reported: \cite{Binder2016} found an age of $<$5 Myr for SN 2010da (an HMXB in NGC 300);  in own galaxy, Cir X-1 has been revealed as the youngest known related example at $<$4600 yr \citep{Heinz2013}.  However, there is currently debate about the nature of the optical companion in SN 2010da (LBV, yellow supergiant, etc.), and the companion in Cir X-1 has proven infamously difficult to classify. \cite{Johnston2016} proposed that the recent SN in Cir X-1 has shock-heated the envelope of the companion, puffing it up, which can reconcile the strangely red color for an HMXB.  Consequently we are wary of excluding red objects from the IC 10 sample and we must obtain spectra of all counterparts. 

In summary, our \chandra~ monitoring study has found a heterogeneous population of exotic HMXBs in IC 10. We are witnessing a starburst at an early stage, even perhaps before its peak \citep{Leroy2006}, so these unfamiliar systems will play an important role in understanding the age/metallicity dependence of massive binaries, the X-ray binary luminosity function, and ULX sources. 

\section{Acknowledgements}

We are most grateful to the anonymous referee for their constructive improvements to the manuscript. This project was made possible by the support of SAO grant NAS8-03060 and the Physics Department of University of Massachusetts, Lowell. SL also thanks the Gemini Observatory for its support. Gemini is operated by the Association of Universities for Research in Astronomy, Inc., on behalf of the international Gemini partnership of Argentina, Australia, Brazil, Canada, Chile, the United Kingdom, and the United States of America.


\end{document}